\documentclass[letterpaper, 10pt, conference]{ieeeconf}

\usepackage{epsfig}
\usepackage{subfigure}
\usepackage{amsmath}
\usepackage{amssymb}
\usepackage[T1]{fontenc}
\usepackage{times}
\fontfamily{ptm}\selectfont

\begin{document}

\title{\LARGE \bf Detection of Markov Random Fields on Two-Dimensional Intersymbol Interference Channels}

\author{\authorblockN{Ying Zhu, Taikun Cheng, Krishnamoorthy Sivakumar, and Benjamin J. Belzer}
\authorblockA{School of Electrical Engineering and Computer Science\\
Washington State University\\
P.O. Box 642752,
Pullman, WA 99164-2752, USA \\
Email: yzhu2,tcheng,siva,belzer@eecs.wsu.edu}}
% avoiding spaces at the end of the author lines is not a problem with
% conference papers because we don't use \thanks or \IEEEmembership
\maketitle
\thispagestyle{empty}
\pagestyle{empty}

\begin{abstract}
We present a novel iterative algorithm
for detection of binary Markov random fields (MRFs) corrupted by 
two-dimensional (2D) intersymbol interference (ISI)
and additive white Gaussian noise (AWGN).  We assume a first-order 
binary MRF
as a simple model for correlated images.
We assume a 2D
digital storage channel, where the MRF is
interleaved before being written and then read by a 2D
transducer; such channels occur in recently proposed 
optical disk storage systems.  The detection algorithm
is a concatenation of two soft-input/soft-output (SISO) detectors:
an iterative row-column soft-decision feedback (IRCSDF) ISI detector,
and a MRF detector.  The MRF detector is a SISO version of
the stochastic relaxation algorithm by Geman and Geman
in IEEE Trans. Pattern Anal. and Mach. Intell., Nov. 1984.
On the $2 \times 2$ averaging-mask ISI channel, at a 
bit error rate (BER) of $10^{-5}$, the
concatenated algorithm achieves SNR savings of between 0.5 and 2.0 dB
over the IRCSDF detector alone; the savings increase as the MRFs become
more correlated, or as the SNR decreases.  The algorithm is also 
fairly robust to mismatches between the assumed and actual MRF parameters. 

\end{abstract}

\section{Introduction}
\begin{figure}[t]
\centerline{\epsfig{figure=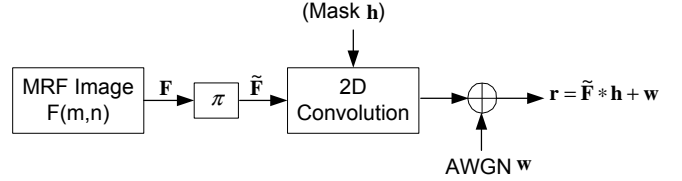, width=3.4in}}
\vspace{-0.1in}
 \caption{Block diagram of the digital storage channel assumed in this paper; the $\pi$ 
represents an image interleaver, and $*$ represents 2D convolution. Level shifting
from $\{0,1\}$ to $\{-1,1\}$, performed after the interleaver, is not shown.}
 \label{fig: cblock}
\end{figure}

This paper considers the detection of an $M \times N$ binary-valued 2D MRF
$F$ transmitted over the digital storage channel shown in Fig.~\ref{fig: cblock}.
The received image
\begin{equation}
r(m,n) = \sum_{k}\sum_{l} h(k,l) \tilde{F}(m-k,n-l) + w(m,n),
\label{eq:2Dconv}
\end{equation}
where $0 \le m \le M-1$, $0 \le n \le N-1$, $h(k,l)$ is a 
finite-impulse-response 2D blurring
mask, $\tilde{F} = 2\times\pi(F)-1$ is an interleaved and level-shifted version of $F$,
the $w(m,n)$
are zero mean i.i.d.\ Gaussian random variables (r.v.s) with variance $\sigma_w^2$, 
and the double sum is computed over
the mask support region $\mbox{$\mathcal S$}_h = \{(k,l):h(k,l) \ne 0 \}$.
The MRF $F$ takes pixel values from $\{0,1\}$, and is correlated according to the following
Markov property \cite{Geman}: 
\begin{equation}
\begin{split}
\mathrm{Pr}(F_{m,n} = f_{m,n} | F_{k,l} = f_{k,l}, (k,l) \ne (m,n)) \\
= \mathrm{Pr}(F_{m,n} = f_{m,n} | F_{k,l} = f_{k,l}, (k,l) \in \mathcal{F}_{m,n}).
\end{split}
\label{eq: Mprop}
\end{equation}
In (\ref{eq: Mprop}), $F_{m,n}$ denotes pixel $(m,n)$ of the MRF,
$f_{k,l}$ denotes a particular value of pixel $(k,l)$, and $\mathcal{F}_{m,n}$
denotes the first-order neighborhood of pixel $(m,n)$: 
$\mathcal{F}_{m,n} = \{(m,n-1),(m,n+1),(m-1,n),(m+1,n)\}$.
The MRF $F$ is generated by a Gibbs sampler based on the Ising model, which is
characterized by a two parameter energy function. 
It is assumed that the detector knows the energy parameters (and therefore the
complete Markov model) of the transmitted MRF.
Because of the interleaver, it is assumed
that the pixels in $\tilde{F}$ are independent and identically distributed 
(i.i.d.), with $\mathrm{Pr}\{F(m,n) = 0\} = p_0 = 1 - p_1$ for all $(m,n)$, where the
{\em a priori} probabilities $p_0$ and $p_1$ need not in general equal $1/2$.

The motivating application for the channel model employed in this paper
is 2D magnetic or optical storage
systems, which are subject to 2D ISI.  Such systems (e.g., \cite{in-phase-mit})
store the image $\tilde{F}$
in its original 2D form, rather than converting the image to a 1D sequence and storing
it on 1D tracks.  Over the past 10 years, a number of
papers (e.g., \cite{HGH96}-\cite{taikun_jour}) have considered the detection problem
for binary images on the 2D ISI channel, under the assumption that the transmitted image
pixels are i.i.d. and equiprobable.  In practice, this i.i.d. assumption can be largely
achieved by interleaving the original image before storage (as in Fig.~\ref{fig: cblock}), 
and deinterleaving it after detection.  However, in practice the original image is 
often correlated.  Uncompressed natural images are highly correlated; these typically 
occur in diagnostic medical
imaging where distortion due to lossy compression is intolerable, and the time required
for lossless compression/decompression may not be available.
Images usually retain some correlation even if stored in compressed format, because practical
lossless (or lossy) image compression schemes do not completely decorrelate the image.

The principal contribution of the present paper is an iterative detection scheme that
exploits the correlation in the original image to achieve significant SNR savings over
ISI detection schemes which do not account for correlated input images.  Although we
model the image correlation by a simple first-order Markov random field, 
our results give at
least a proof of principle that can be further explored with more realistic
image models.
A key innovation introduced in this paper is a soft-input, soft-output MRF 
detection algorithm,
based on the stochastic relaxation image restoration algorithm of \cite{Geman}.  The
ISI detector employed in this paper is described in \cite{taikun,taikun_jour}, where
comparisons with several previously published algorithms show that it is among the best
performing ISI detectors in the public domain.

\section{The Concatenated Detector}
\label{sec: detector}
\begin{figure}[tbh]
\centerline{\epsfig{figure=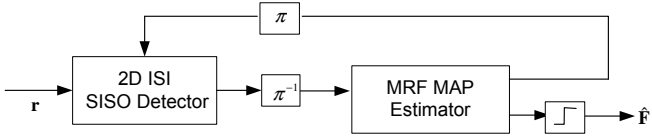, width=3.4in}}
\vspace{-0.1in}
 \caption{Block diagram of the concatenated detector.}
 \label{fig: dblock}
\end{figure}

A block diagram of the concatenated system is shown in Fig.~\ref{fig: dblock}.
It operates according to the ``turbo principle'' (after turbo-codes \cite{Berrou}),
whereby two or more SISO decoders exchange extrinsic information and iterate
until convergence.
The received image $\mathbf{r}$ is an input to the ISI detector, which
attempts to remove the ISI under the assumption that the pixels of $\tilde{F}$
are i.i.d.  The ISI detector outputs deinterleaved extrinsic log-likelihood
ratios (LLRs) to the MRF detector, which estimates the original MRF $F$, and feeds
back interleaved extrinsic LLRs to the ISI detector.  

\subsection{The ISI Detector}
A detailed description of the ISI detector, including performance
comparisons with a number of other previously published 2D ISI
algorithms, appears in \cite{taikun,taikun_jour}.  The ISI detector
is therefore only briefly described here.

\begin{figure}[tbh]
\centerline{\epsfig{figure=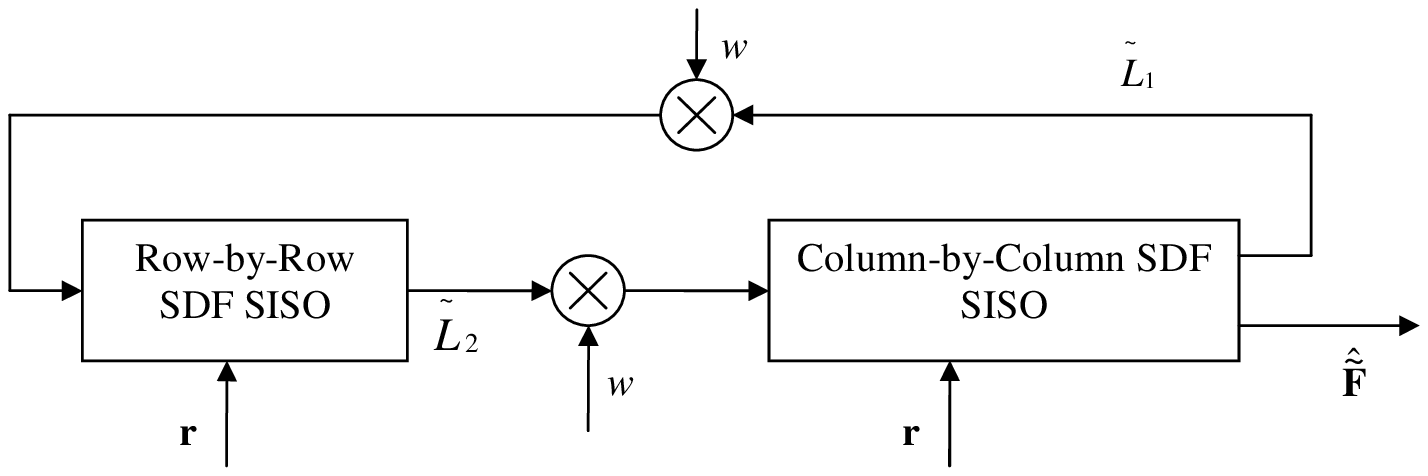, width=3.0in}}
\vspace{-0.1in}
 \caption{ISI detector block diagram.  The extrinsic input from
the MRF detector into the row-SISO is not shown.}
 \label{fig: isiblock}
\end{figure}

Figure~\ref{fig: isiblock} is a block diagram of the ISI detector,
which employs an iterative, row-column soft-decision feedback
algorithm (IRCSDFA).
The IRCSDFA consists of two SISO modules, run on rows and columns, 
which exchange 
weighted soft information estimates
of the interleaved MRF $\tilde{F}$. Each module runs the BCJR algorithm
\cite{bcjr} on several rows (columns) of the image at once, and uses
soft decision feedback from previously-processed rows (columns), to 
arrive at an LLR estimate $\tilde{L}$
of the current row (or column.)  The weight $w$ attenuates the LLR 
estimates, to
correct for the over-confidence effect resulting from use of soft decision
feedback (SDF).  

\begin{figure}[h]
\centerline{\epsfig{figure=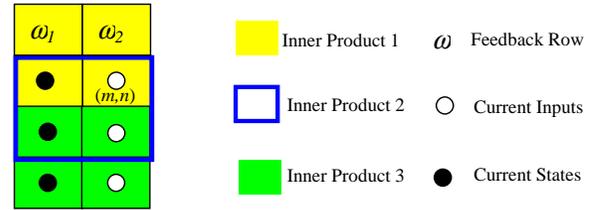, width=3.0in}}
%\vspace{-0.1in}
 \caption{The ISI detector's row-SISO trellis states and inputs, for a $2 \times 2$
mask.}
 \label{fig: isitrell}
\end{figure}

The row-SISO trellis state and input block for the $m$th row of
an image corrupted by $2 \times 2$ ISI is shown 
in Fig.~\ref{fig: isitrell}. A trellis is generated by shifting this
block right one column at
a time through all the pixels in a given row, and taking into account
the eight possible values of the three input pixels.  The resulting trellis
has eight states and eight branches per state.
Each trellis branch metric is the sum
of three terms; each term is the 
squared Euclidean distance (SED) between
one of the 2D inner products shown in Fig.~\ref{fig: isitrell}
and the corresponding received pixel $r(m,n)$, $r(m+1,n)$, or $r(m+2,n)$.
Each 2D inner product is computed between the inverted convolution
mask $h(m-k,n-l)$ and the appropriate block of state, input, and feedback 
pixels indicated
in Fig.~\ref{fig: isitrell}; to compute inner product 2 shown in
the figure, the inverted mask's $h(0,0)$ coefficient overlays
the current pixel position $(m,n)$. 
The algorithm uses the LLRs $\omega_1$ and $\omega_2$ from the previously
processed row as soft-decision
feedback when computing modified pixel transition 
probabilities $\gamma_\mathbf{i}(\mathbf{r}_k,s',s)$ in the BCJR algorithm,
where $\mathbf{i} = [i_0, i_1, i_2]$ and $\mathbf{r} = [r(m,n), r(m+1,n), r(m+2,n)]$ 
denote input and received pixel vectors (as seen in Fig.~\ref{fig: isitrell}),
$k$ denotes the iteration number, and the candidate input pixels $i_j$ 
take values from 
$\{-1, 1\}$ with {\em a priori} probabilities $p_0$ and $p_1$.

The modified $\gamma$ is 
the product of a modified conditional channel PDF $p'(\cdot)$ (which takes into
account the SDF LLRS $\omega_1$ and $\omega_2$, and which is described further
in \cite{taikun,taikun_jour}), trellis transition 
probabilities, and
extrinsic information from the other
decoder:
\begin{equation}
\begin{split}
&\gamma_\mathbf{i}(\mathbf{r}_k,s',s)=p'(\mathbf{r}_k|\mathbf{u}=\mathbf{i},S_k=s,S_{k-1}=s')\\
&\times P(\mathbf{u}=\mathbf{i}|s,s')\times P(s|s')\times P(\mathbf{u}=\mathbf{i}|\tilde{L}).
\label{mod_gamma}
\end{split}
\end{equation}
In (\ref{mod_gamma}),
$\mathbf{u} = [u_0, u_1, u_2]$ denotes estimate of the input vector
$[\tilde{F}(m,n), \tilde{F}(m+1,n), \tilde{F}(m+2,n)]$.
For the given states $ s',s $ and input $\textbf{u}$, $ P(\mathbf{u}=\mathbf{i}|s,s') $ is 0 or 1. Because the current input vector becomes the next state
vector, the state transition probability is the product of the {\em a priori}
probabilities ($p_0$ or $p_1$) of the independent components of $\mathbf{u}$: $ P(S_k=s|S_{k-1}=s') = \prod_{j=0}^{3}P(u_j = i_j)$. The extrinsic information probability $P(\mathbf{u}=\mathbf{i}|\tilde{L})$ is derived 
from the
extrinsic input LLRs from the other detector.

The above-described IRCSDFA is similar to the ISI algorithm described
in \cite{marrowconf03,marrowvgs}, but with a number of key differences:
(1) we have adapted our algorithm to handle non-equiprobable binary images;
(2) we make decisions on only one row at a time, as opposed to multiple 
rows; (3) we use three inner products for a $2 \times 2$ mask, instead of
two; and (4) we weight the LLRs passed between SISO modules.  We have 
observed experimentally that the last two differences improve
the performance of the IRCSDFA by about 1 dB, when it is run
on source images of non-trivial size (i.e., $64 \times 64$ or bigger).

To the best of our knowledge,
the above-described IRCSDFA gives the best published performance for
equalization of the $2 \times 2$ averaging mask on 
non-trivially-sized source images, 
with the 
exception of the concatenated zig-zag algorithm 
of \cite{patrick}, which uses the IRCSDFA as a component.  
We chose the IRCSDFA for the experiments described in this paper
because it is
about half as complex as the concatenated zig-zag algorithm, and because
the focus in this paper is on the improvements provided by concatenating
a MRF detector with an ISI detector.

\subsection{The MRF Detector}
\label{subsec: MRF}
The MRF detector is designed to provide a maximum {\em a posteriori} (MAP) 
estimate of the original MRF $F$ from
its noisy version $G = F+z$, where $z$ is zero mean 2D AWGN.
The MRF detector's extrinsic input LLRs are the deinterleaved extrinsic
output LLRs from the ISI detector: 
\begin{displaymath}
L_\mathrm{in}^\mathrm{MRF} = 
\pi^{-1}\left(L_{\mathrm{out}}^{\mathrm{ISI}}\right).
\end{displaymath}
In practice, the $L_\mathrm{in}^\mathrm{MRF}$
are (approximately) conditionally normal, with conditional means 
$\mu_+$ and $\mu_-$ corresponding to pixel values of $+1$ or $-1$ 
in $\pi^{-1}(\tilde{F})$.
The MRF detector computes sample mean and variance estimates $\hat{\mu}_+$,
$\hat{\mu}_-$, $\hat{\sigma}^2_+$, and $\hat{\sigma}^2_-$ for the two conditional
input PDFs.  The LLRs $L_\mathrm{in}^\mathrm{MRF}$
are then shifted and scaled to form the ``noisy image'' $G$, which has conditional
means of 0 and 1:
\begin{equation} 
G(m,n) = (L_\mathrm{in}^\mathrm{MRF}(m,n) - \hat{\mu}_-)/(\hat{\mu}_+ - \hat{\mu}_-).
\end{equation}
The conditional variances of $G$ are 
estimated as 
\begin{equation}
\sigma_G^2 =  (N_+ \hat{\sigma}^2_+ + N_- \hat{\sigma}^2_-)/((N_+ + N_-)(\hat{\mu}_+ - \hat{\mu}_-)^2), 
\label{eq: sig2G}
\end{equation}
where $N_+$ and $N_-$ are the 
number of positive and negative pixels
in the input LLR image $L_\mathrm{in}^\mathrm{MRF}$.

\subsubsection{Generation of the MRFs}
The conditional probabilities in (\ref{eq: Mprop}) are
calculated according to the Gibbs distribution \cite{Geman}
\begin{equation}
\begin{split}
\mathrm{Pr}(F_{m,n} = f_{m,n} | F_{k,l} = f_{k,l}, (k,l) \in \mathcal{F}_{m,n}) \\ 
= \frac{e^{-\mathcal{E}(f_{m,n})/T}}{\sum_{f=0}^1 e^{-\mathcal{E}(f)/T}}.
\end{split}
\label{eq: prob}
\end{equation}
The energy function $\mathcal{E}$ used to generate the MRFs in this paper 
follows the Ising model:
\begin{equation}
\mathcal{E}_I(f_{m,n}) = f_{m,n}(\alpha + \beta v_{m,n}),
\label{ising}
\end{equation}
where $v_{m,n} = f_{m,n-1}+f_{m,n+1}+f_{m-1,n}+f_{m+1,n}$.  The MRF becomes more
correlated as the interaction coefficient $\beta$ becomes increasingly large
and negative.
Coefficient
$\alpha$ is related to the prior probability of pixel $(m,n)$; $\alpha$ is set
equal to $-2\beta$ when the pixels are equiprobable.
The ``temperature'' parameter $T$ is
set to one to generate the original MRFs, but is varied according to an 
annealing schedule
during the stochastic relaxation algorithm used for MRF estimation.   

The method used to simulate the MRF is as follows\cite{Cross}:

\begin{enumerate}
\item Start with an i.i.d random configuration.

\item Randomly chose two pixels.

\item Compute the energy change $\Delta\mathcal{E}$ if these two pixels are switched.

\item If $\Delta\mathcal{E} < 0$, i.e., if the energy decreases, accept the switch.

\item Otherwise, accept the switch with probability $q \propto \exp \left(-\Delta\mathcal{E}\right)$.

\item Go to 2) until the convergence occurs.
\end{enumerate}

\begin{figure}[htbp]
  \begin{center}
    \mbox{
      \subfigure[]{\scalebox{0.2}{\epsfbox{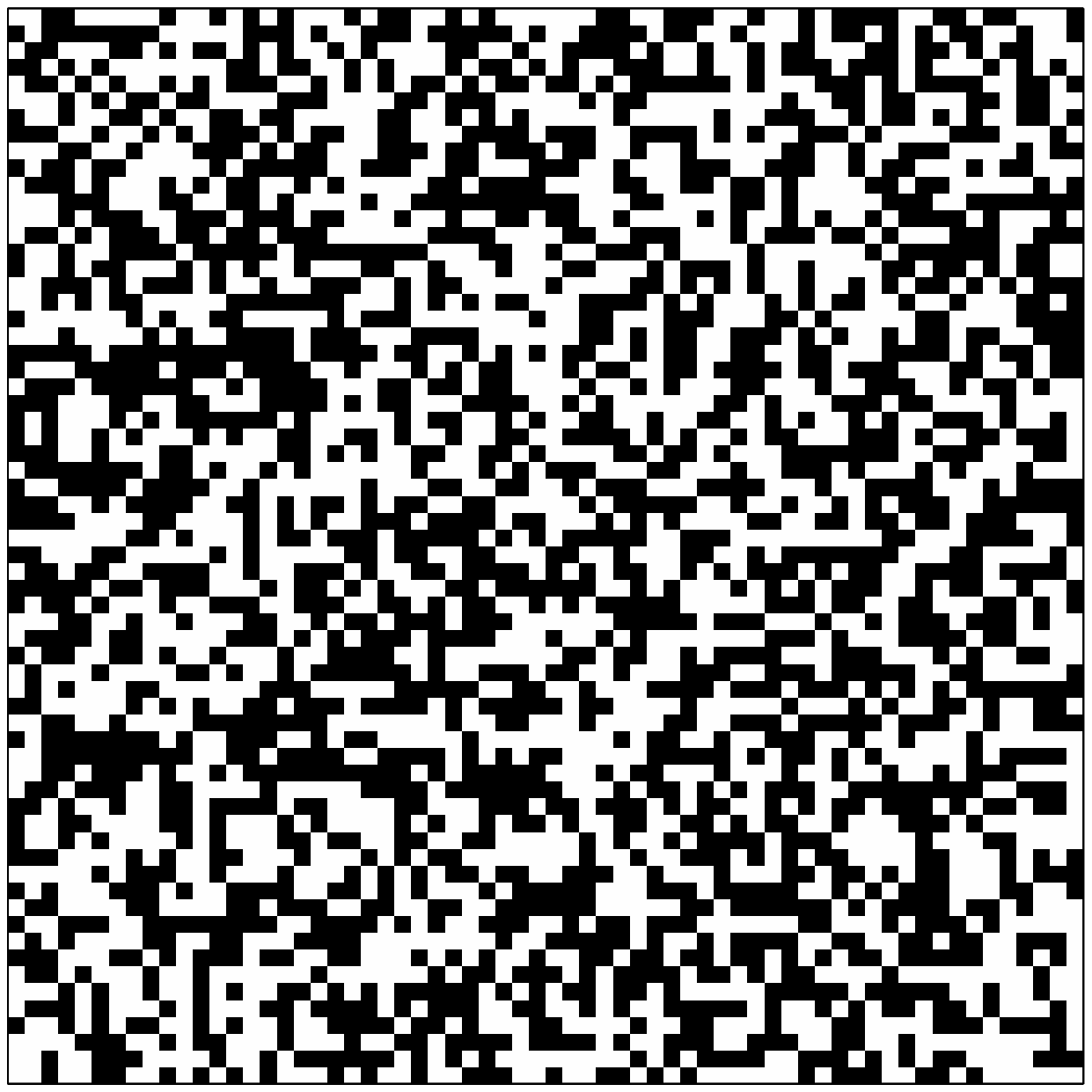}}} \quad
      \subfigure[]{\scalebox{0.2}{\epsfbox{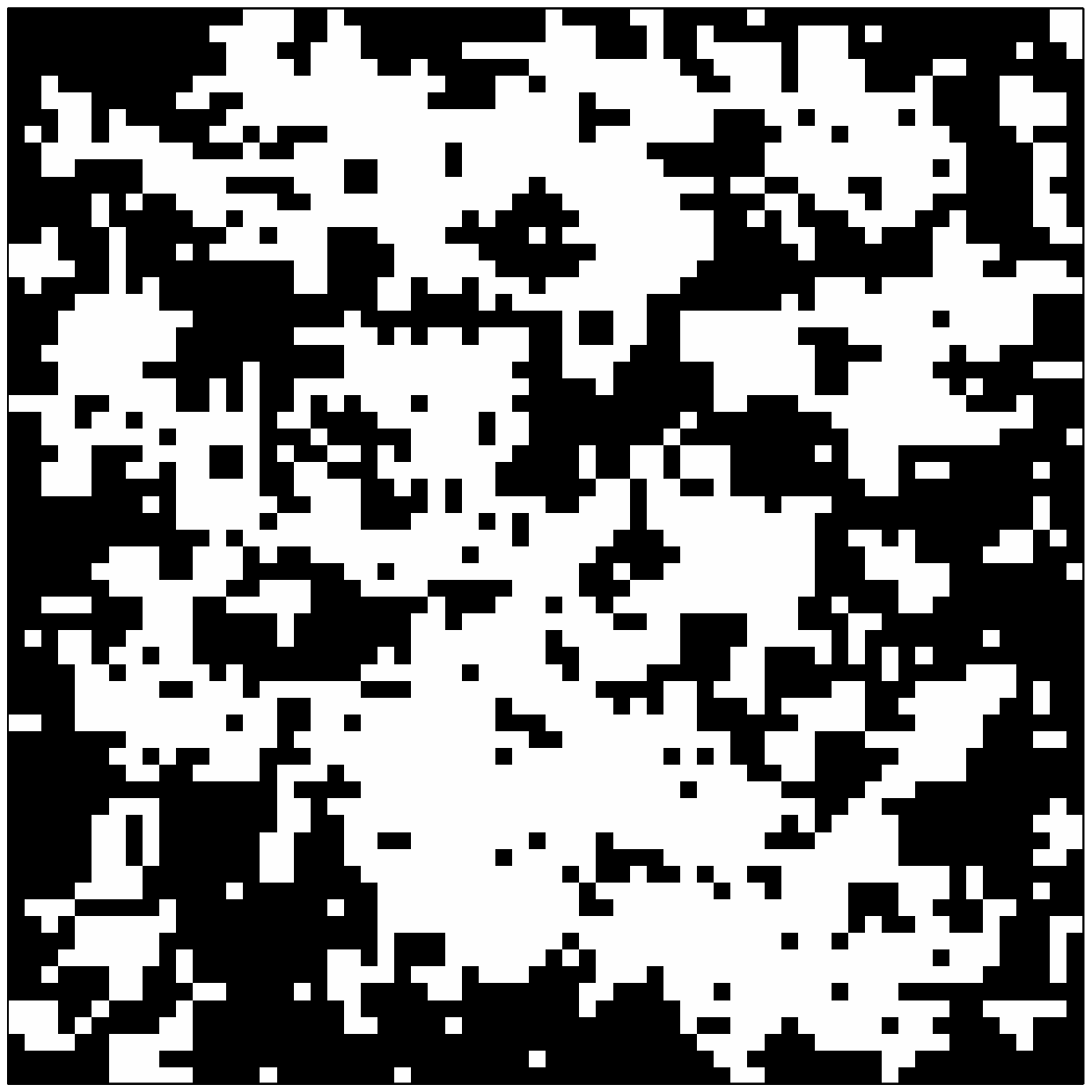}}} \quad
      \subfigure[]{\scalebox{0.2}{\epsfbox{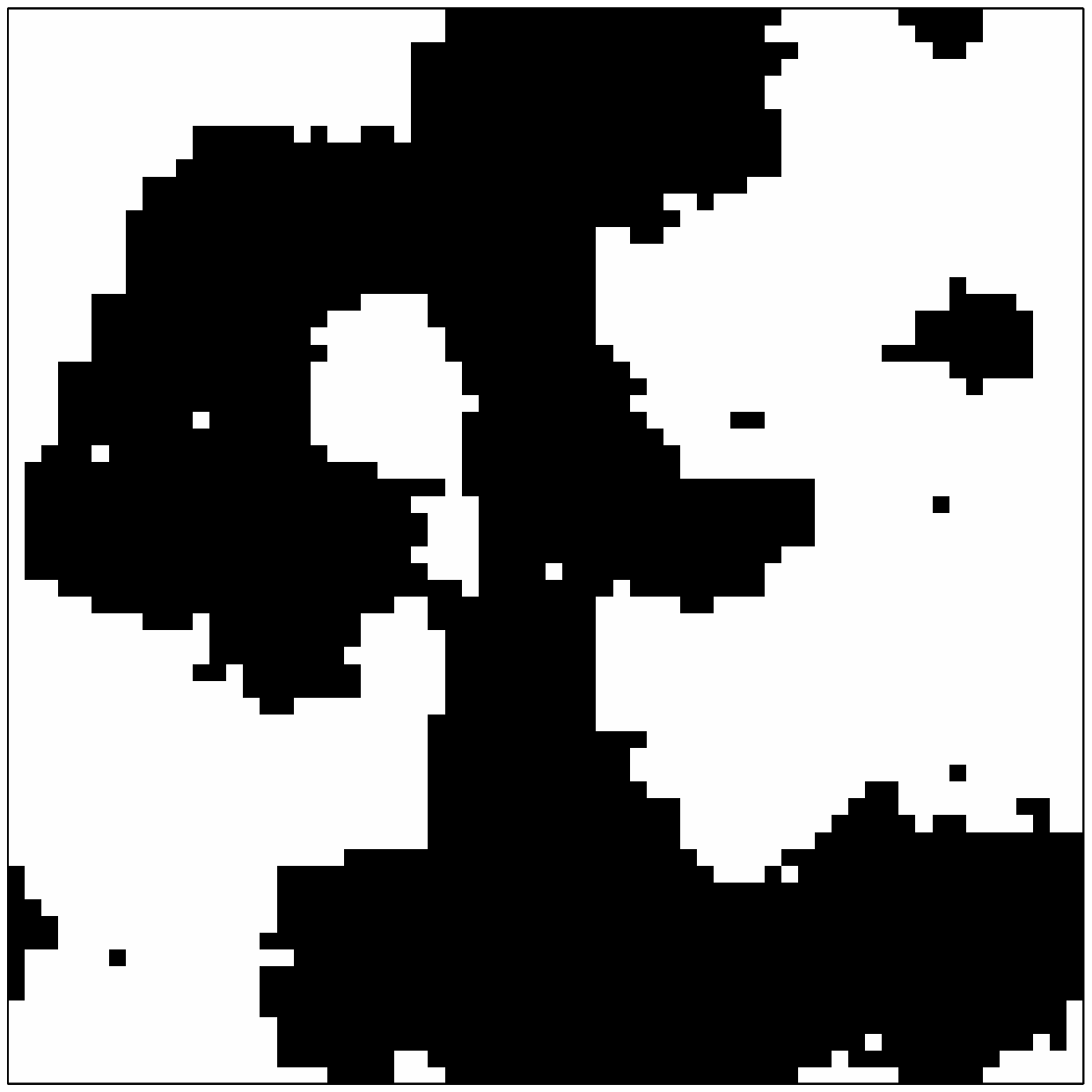}}}
      }
    \caption{Three $64 \times 64$ equiprobable binary images: (a) i.i.d. image; (b) MRF with correlation parameter $\beta=-1.5$; (c) MRF with $\beta=-3.0$.}
    \label{fig: ims}
  \end{center}
\end{figure}

This method ensures that the generated MRF has the same 
number of 0s and 1s as the initial configuration,
and after the interleaver it has the same distribution as an i.i.d.
 source. Two examples of generated equiprobable MRFs are shown as images
b and c in Fig.~\ref{fig: ims}.  In these images, black represents 0, and white represents 1.
Image b has $\beta=-1.5$, and image c has $\beta=-3.0$; from the figure
it is clear that image c is more correlated than image b.  Fig.~\ref{fig: ims} (a) shows an i.i.d. image for comparison.

\begin{figure}[htb]
\centerline{\epsfig{figure=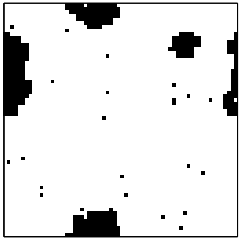, width=2.0in}}
 \caption{Non-equiprobable $64 \times 64$ binary MRF, with $p_0 = 0.1$ and $\beta = -3.0$.}
 \label{fig: prob}
\end{figure}

If the source image is non-equiprobable, then 
parameter $\alpha$ in the MRF model needs to be modified as follows.
If we level shift the binary alphabet for $F_{m,n}$ such that $0 \rightarrow -1$ and
$1 \rightarrow 1$, let $\beta=0$, and denote the level-shifted pixels by $f_{m,n}'$,
then $\mathcal{E}_I(f_{m,n}')=f_{m,n}'\alpha_1$, where we use the
notation $\alpha_1$ to denote the alpha constant for the $\{-1, 1\}$
alphabet.
So, 
\begin{equation*}
p_0 \equiv P(F_{m,n}'=-1) = \frac{e^{\alpha_1}}{e^{-\alpha_1}+e^{\alpha_1}}
\end{equation*}
\begin{equation*}
p_1 \equiv P(F_{m,n}'=+1) = \frac{e^{-\alpha_1}}{e^{-\alpha_1}+e^{\alpha_1}},
\end{equation*}
which gives
\begin{equation*}
\alpha_1 = \frac{1}{2}\log\frac{P(F_{m,n}'=-1)}{P(F_{m,n}'=+1)} = 
\frac{1}{2}\log\frac{p_0}{p_1}.
\end{equation*}
Now, we map the values from $\{-1, 1\}$ to $\{0, 1\}$ and consider the energy
function in (\ref{ising}):
\begin{eqnarray*}
    \mathcal{E}_I(f_{m,n})    
     &= & f_{m,n}'(\alpha_1 + \beta v_{m,n}') \\
     &= & (2f_{m,n}-1) \{\alpha_1 + \beta [ (2f_{m,n-1}-1) \\
     &  & + (2f_{m,n+1}-1)+(2f_{m-1,n}-1) \\ 
     &  & + (2f_{m+1,n}-1) ] \} \\
     &= & 2\alpha_1 f_{m,n} + 4\beta v_{m,n} f_{m,n} -8\beta f_{m,n} \\
     &  & - 2\beta v_{m,n} + 4\beta - \alpha_1 \\
     &= & \frac{1}{4} f_{m,n} (\frac{1}{2}\alpha_1 - 2\beta + \beta v_{m,n}) \\
     &  & - 2\beta v_{m,n} + 4\beta - \alpha_1 \\
     &= & \frac{1}{4} f_{m,n} (\alpha + \beta v_{m,n}) - 2\beta v_{m,n} + 4\beta - \alpha_1,
\end{eqnarray*}
where 
\begin{equation}
\alpha = \frac{1}{2}\alpha_1 - 2\beta = \frac{1}{4}\log\frac{p_0}{p_1} - 2\beta.
\label{eqn: alpha_nep}
\end{equation}
Given the neighbors $v_{m,n}, $$- 2\beta v_{m,n} + 4\beta - \alpha_1 $ is a constant 
which does not affect the probability computation.
An example MRF with $(p_0, p_1) = (0.1, 0.9)$ is shown in Fig.~\ref{fig: prob}.

\subsubsection{Stochastic Relaxation}
The stochastic relaxation algorithm of Geman and Geman (G\&G) \cite{Geman} is an iterative algorithm 
that proceeds
at discrete time steps $t=0,1,2,\ldots,t_{\mathrm{max}}$; after each step, a new 
MRF estimate $\hat{F}(t)$ is obtained.  For sufficiently large $t_{\mathrm{max}}$,
the algorithm converges to a final estimate $\hat{F}$ that does not change appreciably
for $t > t_{\mathrm{max}}$. 
The initial
estimate $\hat{F}(0)$ is computed by thresholding the noisy image $G$
at $1/2$.  Given estimate $\hat{F}(t)$ at time $t$,
at time $t+1$ $M \times N$ randomly chosen pixels of $\hat{F}(t)$ are visited.
The value of each pixel $(m,n)$ visited during the random scan is set to
0 or 1 with probability $1/2$. If the
new value is different from 
its value $\hat{f}_{m,n}$ at time $t$, then 
the energy difference $\Delta\mathcal{E}_P(m,n) =
\mathcal{E}_P(\mathrm{NOT}(\hat{f}_{m,n}))-\mathcal{E}_P(\hat{f}_{m,n})$ is computed.
In 
computing $\Delta\mathcal{E}_P(m,n)$, only one pixel at a time (i.e., pixel $(m,n)$) is changed;
all other pixels retain their values from time $t$.
If $\Delta\mathcal{E}_P(m,n) < 0$, then the change is 
accepted: $\hat{f}_{m,n}(t+1) = \mathrm{NOT}(\hat{f}_{m,n}(t))$.  If $\Delta\mathcal{E}_P(m,n) \ge 0$,
the change is accepted with probability $q = \exp\left[-\Delta\mathcal{E}_P(m,n)/T(t+1)\right]$.
The temperature $T$ is gradually reduced according to a logarithmic annealing schedule:
$T(t) = C/\log(1+t), 1 \le t \le t_{\mathrm{max}}$; in this paper the value $C = 3.0$
is used for all simulations. 
 
The modified ({\em a posteriori}) energy 
function $\mathcal{E}_P$ used by the MRF estimator at time $t+1$
includes the difference energy between the noisy input image $G$ and the 
current trial estimate:
\begin{equation}
\mathcal{E}_P(f_{m,n}) = \mathcal{E}_I(f_{m,n}) + \|G - \hat{F}(t,f_{m,n})\|^2/2\sigma_G^2,
\label{eq: E_P}
\end{equation}
where $\hat{F}(t,f_{m,n})$ denotes the estimated MRF $\hat{F}(t)$ with pixel $(m,n)$ 
taking the value $f_{m,n}$, and $\sigma_G^2$ is estimated as in (\ref{eq: sig2G}).
The extrinsic information LLRs $L_\mathrm{in}^\mathrm{MRF}(m,n) = 
\log\left[\mathrm{Pr}_\mathrm{ext}(F_{m,n} = 1)/\mathrm{Pr}_\mathrm{ext}(F_{m,n} = 0)\right]$ represent 
independent {\em a priori} 
information
about pixel $(m,n)$. Hence, in the $\mathcal{E}_I(f_{m,n})$ of (\ref{eq: E_P}), 
we replace $\alpha$ with 
\begin{equation}
\alpha' = \alpha -L_\mathrm{in}^\mathrm{MRF}(m,n).
\end{equation}
After converting to probabilities
using (\ref{eq: prob}), this extra LLR term in $\alpha'$
results in the corresponding extrinsic probabilities 
$\mathrm{Pr}_\mathrm{ext}(F_{m,n} = f_{m,n})$ appearing as independent weight factors in 
the renormalized expressions
for the conditional probabilities of (\ref{eq: prob}). 
Thus, as the LLRs $L_\mathrm{in}^\mathrm{MRF}(m,n)$ grow increasingly large with
successive iterations of the concatenated detector, they
increasingly influence the estimates $\hat{F}$ arrived at by the MRF detector.  

The G\&G algorithm provides only a binary estimate $\hat{F}$ of the MRF.
To compute LLR estimates $L^\mathrm{MRF}(m,n)$ for each pixel, we use the
following method, which is similar to that
in \cite{mertins1,mertins2}, where a soft-output
G\&G algorithm is used
for recovery of noise-corrupted MRFs, and for iterative
source-channel image decoding with MRF source models. 
After the stochastic relaxation algorithm convergences, 
we compute the conditional probabilities based on the MRF model:
\begin{equation} 
\mathrm{Pr}(F_{m,n}=0 |\mathcal{F}_{m,n}) = \frac{1}{1+ e^{-(\alpha ' + \beta v_{m,n})/T}}
\end{equation}
\begin{equation} 
\mathrm{Pr}(F_{m,n}=1 |\mathcal{F}_{m,n}) = \frac{ e^{-(\alpha ' + \beta v_{m,n})/T}}{1+ e^{-(\alpha ' + \beta v_{m,n})/T}}.
\end{equation}
The LLRs $L^\mathrm{MRF}(m,n)$ are then computed as
\begin{eqnarray}
\hspace{-0.2in}L^\mathrm{MRF}(m,n) & = & -(\alpha' + \beta v_{m,n})/T \nonumber \\
       & = & -\frac{(\alpha - L_\mathrm{in}^\mathrm{MRF}(m,n) + \beta v_{m,n})}{T}.
\end{eqnarray}
And since $L_\mathrm{in}^\mathrm{MRF}(m,n)$ is weighed by $1/T$, 
\begin{eqnarray}
L_\mathrm{out}^\mathrm{MRF}(m,n) &=&  L^\mathrm{MRF}(m,n) - L_\mathrm{in}^\mathrm{MRF}(m,n)/T \nonumber \\
& = & -(\alpha + \beta v_{m,n})/T. 
\end{eqnarray}

\section{Simulation Results}
\label{sec: sim}

\begin{figure}[tbh]
\centerline{\epsfig{figure=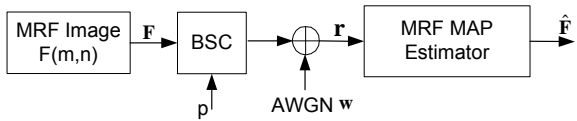, width=3.0in}}
\vspace{-0.1in}
 \caption{Experiment to test error correction capability of the MRF detector.}
 \label{fig: mrfbsc}
\end{figure}

\begin{figure}[tbh]
\centerline{\epsfig{figure=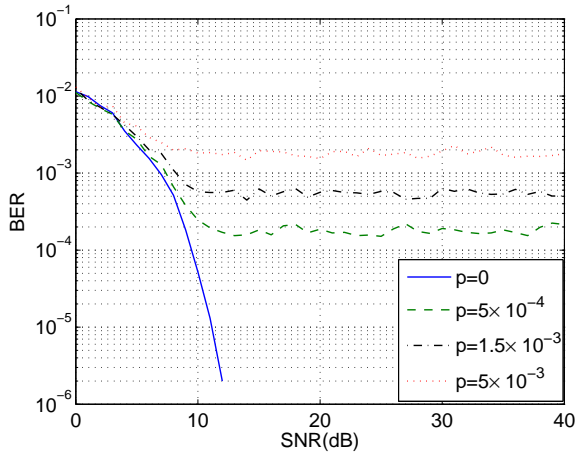, width=3.3in}}
\vspace{-0.1in}
 \caption{Simulation results for the experiment shown in Fig.~\ref{fig: mrfbsc}.}
 \label{fig: bscresults}
\end{figure}

In early iterations of the concatenated system of Fig.~\ref{fig: dblock}, the 
ISI detector cannot remove all the ISI from 
received image $r$.  Due to its use of soft-decision feedback, the
ISI detector is also subject to error propagation, especially at low SNRs.
Thus, the ``noisy image'' $G$ supplied to the MRF detector by the ISI detector contains
bit errors as well as Gaussian-like noise.  To verify that the MRF detector can correct some of these
bit errors (by exploiting the Markov structure of the image), we performed the simulation diagrammed
in Fig.~\ref{fig: mrfbsc}.  The MRF passes through a binary-symmetric channel (BSC) with crossover
probability $p$, followed by an AWGN channel, and is then detected with the MRF MAP estimator.
Fig.~\ref{fig: bscresults} plots the BER of the hard decisions made at the MRF estimator's output versus
the SNR $10\log_{10}\left(\mbox{var}\left[F\right]/\sigma_w^2\right)$ of the AWGN channel, for several values of the BSC error probability $p$.  
In every case,
the MRF detector's output has an error floor lower than the 
value of $p$ used in the simulation.  This result strongly suggests that the MRF detector can improve
the reliability of the information passed to it by the ISI detector.  

All simulations of the concatenated system of Fig.~\ref{fig: dblock} 
used the the $2 \times 2$ averaging mask (with $h(k,l) = 1/4$ for $0 \le (k,l) \le 1$)
in the 2D convolution of (\ref{eq:2Dconv}).  All simulations used 5 outer iterations
of the entire concatenated system, with one inner iteration of the ISI detector performed for each
outer iteration.  In the following subsections, we present simulation results for three cases:
(1) the source image is a binary first order MRF, and the receiver knows the MRF parameters;
(2) the source image is a binary first order MRF, and the receiver guesses the MRF parameters; and
(3) the source image is a natural binary image, and the receiver guesses the most appropriate
MRF parameters. 

\subsection{MRF Source and Known Markov Parameters at the Receiver}
\begin{figure}[tbh]
\centerline{\epsfig{figure=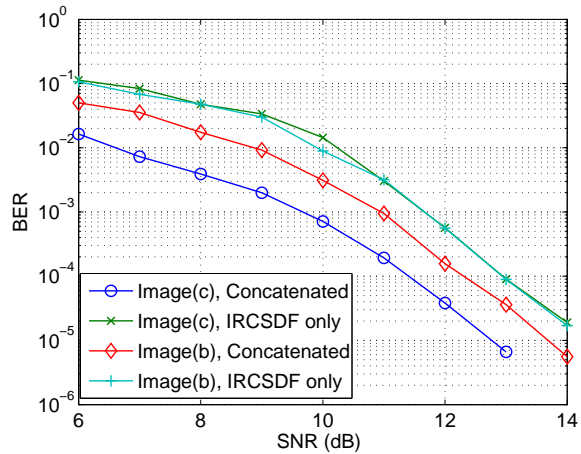, width=3.3in}}
\vspace{-0.1in}
 \caption{Simulation results for the concatenated system of Fig.~\ref{fig: dblock}
on the 2D ISI channel with
$2 \times 2$ averaging mask and AWGN, for MRFs b and c shown in Fig.~\ref{fig: ims}.
The performance of the ISI detector alone is also shown for comparison.}
 \label{fig: isiresults}
\end{figure}

Monte-Carlo simulation results for the concatenated system, when the source images were
the two $64 \times 64$ binary equiprobable MRFs 
shown in Fig.~\ref{fig: ims}(c) and (d), are shown 
in Fig.~\ref{fig: isiresults}.  
The SNR in Fig.~\ref{fig: isiresults} is defined as in \cite{taikun}:
\begin{equation}
\mbox{SNR} = 10\log_{10}\left(\mbox{var}\left[\tilde{F} \ast h\right]/\sigma_w^2\right),
\label{eq:SNR}
\end{equation}
where $\ast$ denotes 2D convolution,
and $\sigma_w^2$ is the variance of the Gaussian r.v.s $w(m,n)$ in~(\ref{eq:2Dconv}).
The performance
of the IRCSDF ISI detector alone on the received image $\mathbf{r}$ is also shown for comparison.  
When the BER is $2 \times 10^{-5}$, 
the concatenated system gives SNR savings of $0.5$ and $1.5$ dB over the ISI detector alone,
for images b and c respectively.
As the input MRF becomes more correlated, the MRF detector makes 
increasingly reliable decisions, thereby improving the system's SNR gain.
At BER of $4 \times 10^{-3}$, the concatenated system saves about $1.0$ and $2.7$ dB 
over the ISI detector alone,
for images b and c respectively. 
The gains increase at lower SNR, where the additional input of the MRF detector 
helps the ISI detector resolve an increased number of ambiguous cases.  

\begin{figure}[tbh]
\centerline{\epsfig{figure=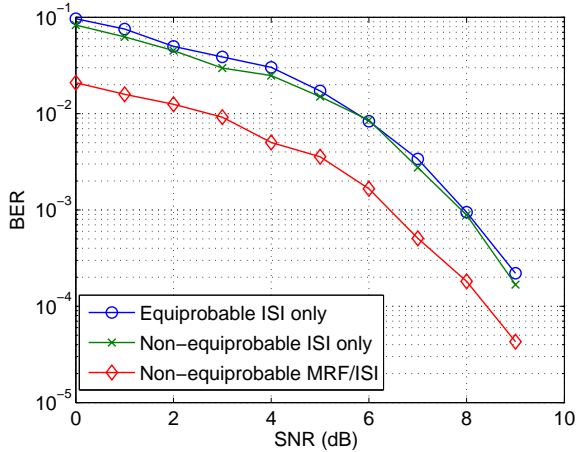, width=3.3in}}
\vspace{-0.1in}
 \caption{Simulation results for the concatenated system of Fig.~\ref{fig: dblock}
on the 2D ISI channel with
$2 \times 2$ averaging mask and AWGN, for the non-equiprobable MRF with $p_0 = 0.1$ and 
$\beta = -3.0$ shown in Fig.~\ref{fig: prob}.
The performances of the equiprobable and non-equiprobable ISI detectors alone are also shown for 
comparison.}
 \label{fig: 0p9isiresults}
\end{figure}

\begin{figure}[tbh]
\centerline{\epsfig{figure=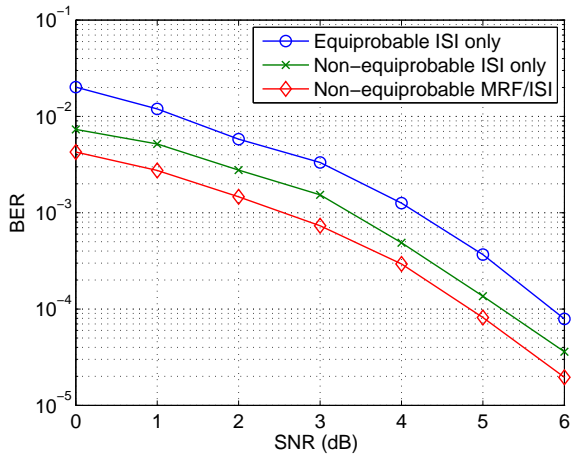, width=3.3in}}
\vspace{-0.1in}
 \caption{Simulation results for the concatenated system of Fig.~\ref{fig: dblock}
on the 2D ISI channel with
$2 \times 2$ averaging mask and AWGN, for a non-equiprobable MRF with $p_0 = 0.01$ and 
$\beta = -3.0$.
The performances of the equiprobable and non-equiprobable ISI detectors alone are also shown for 
comparison.}
 \label{fig: 0p99isiresults}
\end{figure}

We also considered first order MRFs with non-equiprobable pixels.  Fig.~\ref{fig: 0p9isiresults}  
shows simulation results on the $p_0 = 0.1$, $\beta = -3.0$ MRF of Fig.~\ref{fig: prob}.  The
figure shows performance of the concatenated system, as well as the performances of the
non-equiprobable and equiprobable ISI detectors alone.  Non-equiprobable ISI detection
offers little gain at high SNR, but saves between 0.5 and 1 dB at lower SNRs.  
The addition of the MRF detector (with modified $\alpha$ as in (\ref{eqn: alpha_nep})) gives
SNR savings of about 1 dB at BER $2 \times 10^{-4}$, and about 3 dB at BER $10^{-2}$, over
non-equiprobable ISI detection alone.  A similar set of simulation results, but with $p_0 = 0.01$,
is shown in Fig.~\ref{fig: 0p99isiresults}.  Now the gain of non-equiprobable ISI compared to
equiprobable ISI detection is higher: about 0.5 dB at BER $10^{-4}$.  But the gain of the concatenated
system over non-equiprobable ISI is smaller: about 0.3 dB at BER $2 \times 10^{-4}$.  Clearly it
is worth exploiting a non-uniform pixel distribution by using the non-equiprobable concatenated
system in place of equiprobable ISI detection alone, although for extremely skewed
distributions, 
non-equiprobable ISI detection alone achieves most of
the available SNR gain.     

\begin{figure}[tbh]
\centerline{\epsfig{figure=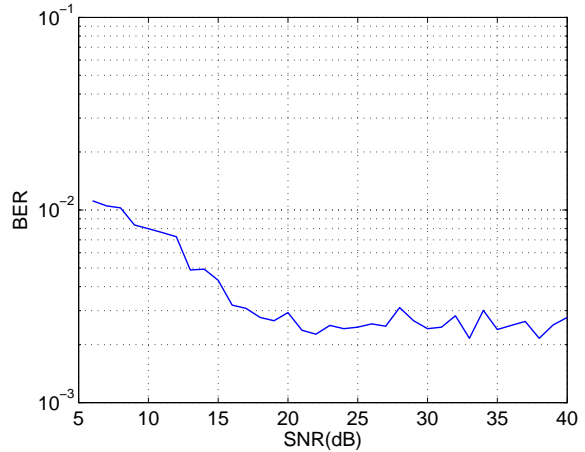, width=3.3in}}
\vspace{-0.1in}
 \caption{Simulation results for the standard Geman and Geman stochastic relaxation
algorithm applied to a MRF with correlation parameter $\beta=-3.0$, that has 
passed through the 2D ISI channel with $2 \times 2$ averaging mask and AWGN without
being interleaved first.}  
 \label{fig: GGalone}
\end{figure}

It is also worth noting that the concatenated algorithm greatly outperforms the 
standard G\&G algorithm applied to a non-interleaved MRF passed through the ISI channel.
Simulation results for the G\&G algorithm alone operating on a $\beta=-3.0$ MRF passed
through the $2 \times 2$ averaging-mask ISI channel are shown in Fig.~\ref{fig: GGalone}.  The SNR in this figure is defined by replacing $\tilde{F}$ by
$F$ in (\ref{eq:SNR}).
From the figure it is clear that the G\&G algorithm alone has a BER floor
of about $1.5 \times 10^{-3}$, and thus is not suitable for 2D digital storage applications, where
much lower BERs are required.

\subsection{MRF Source and Unknown Markov Parameters at the Receiver}
If the MRF detector must estimate the Markov model parameters, 
there will be some estimation error between the actual and estimated parameters.
This leads to the question:
how accurate must the model parameters be in order to achieve performance gains
similar to those seen when the parameters are known exactly?

\begin{figure}[b]
\centerline{\epsfig{figure=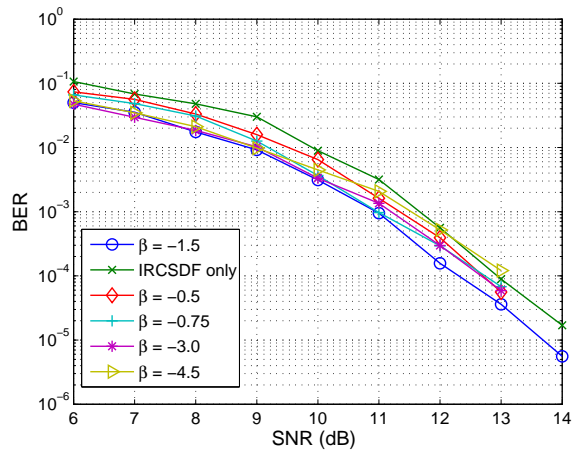, width=3.3in}}
\vspace{-0.1in}
\caption{Simulation results for the concatenated system of Fig.~\ref{fig: dblock}
on the 2D ISI channel with
$2 \times 2$ averaging mask and AWGN, for the $\beta = -1.5$ binary MRF of 
Fig.~\ref{fig: ims}(b) with receiver parameter mismatch.}
\label{fig: mismatcha}
\end{figure}

\begin{figure}[htbp]
\centerline{\epsfig{figure=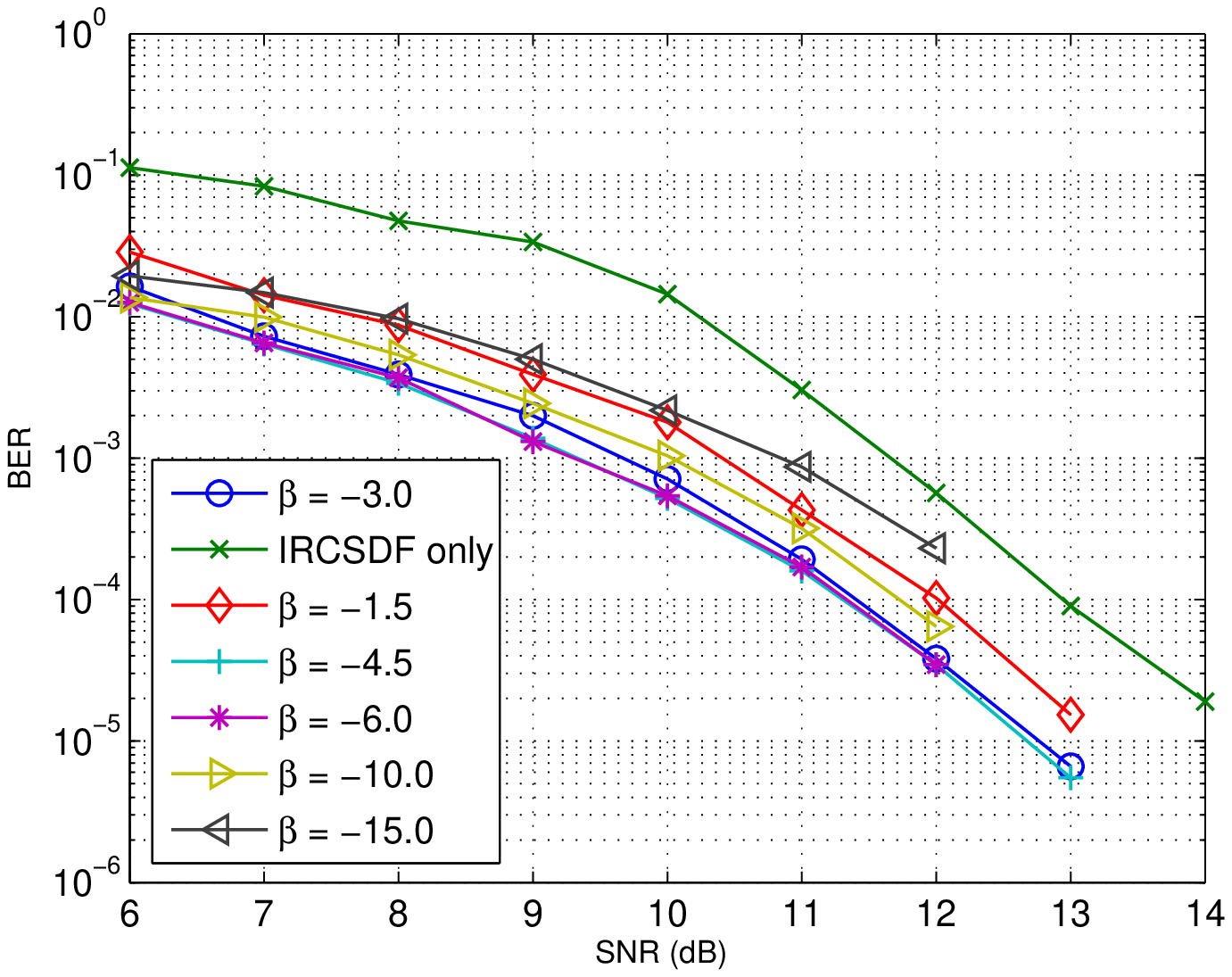, width=3.3in}}
\vspace{-0.1in}
\caption{Simulation results for the concatenated system of Fig.~\ref{fig: dblock}
on the 2D ISI channel with
$2 \times 2$ averaging mask and AWGN, for the $\beta = -3.0$ binary MRF of 
Fig.~\ref{fig: ims}(c) with receiver parameter mismatch.}
\label{fig: mismatchb}
\end{figure}

To answer this question, we did a number of simulations using the equiprobable MRFs
of Fig.~\ref{fig: ims}(b) and (c), in which the assumed MRF parameter $\beta$ did not match
the actual parameter.   These simulation results are shown in Figs.~\ref{fig: mismatcha} and
\ref{fig: mismatchb}.  In Fig.~\ref{fig: mismatcha} the source image has $\beta=-1.5$.
At high SNR, there is a mismatch penalty
of about 0.2 or 0.3 dB when the receiver assumes that $\beta$ is one of
the values $(-0.5,-0.75,-3.0)$, but this still leads to a gain of about 0.2 dB over the ISI-only
case.  Only when $\beta$ is assumed to be $-4.5$ does the concatenated system perform worse
than the ISI detector alone.  Interestingly, at low SNR, the assumed values of $-3.0$ and $-4.5$
give almost identical results to the correct value of $-1.5$.  The concatenated system
thus appears to be quite robust to receiver parameter mismatch, at both high and low SNRs. 

The results shown in Fig.~\ref{fig: mismatchb}, where the source MRF has $\beta=-3.0$, suggest
that robustness to parameter mismatch increases as the source MRF becomes more correlated.  
At high SNR, assumed values of $\beta$ in a range between $-1.5$ and $-10.0$ give gains of 
more than 1 dB compared to ISI-only detection; assumed values of $-4.5$ and $-6.0$ perform as
well as the correct value of $\beta=-3.0$.

\subsection{Natural Image Sources}
\begin{figure}[htbp]
  \begin{center}
    \mbox{
      \subfigure[]{\scalebox{0.25}{\epsfbox{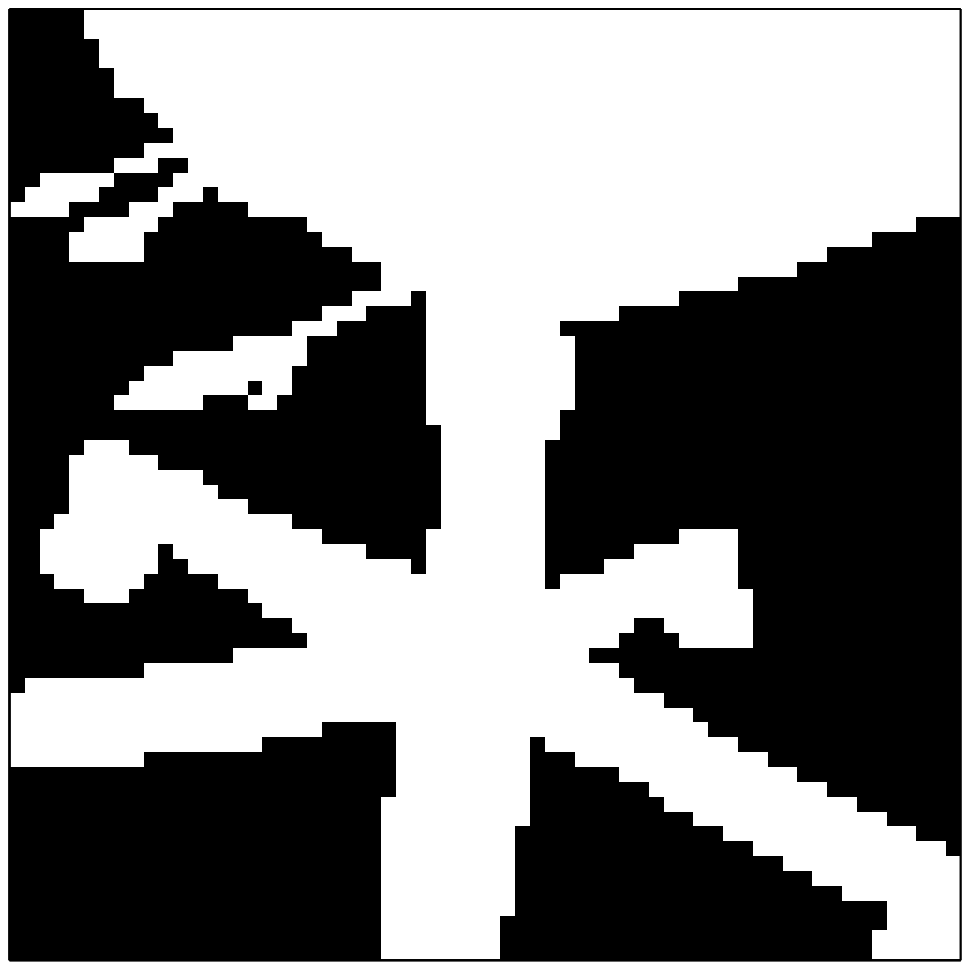}}} \quad
      \subfigure[]{\scalebox{0.25}{\epsfbox{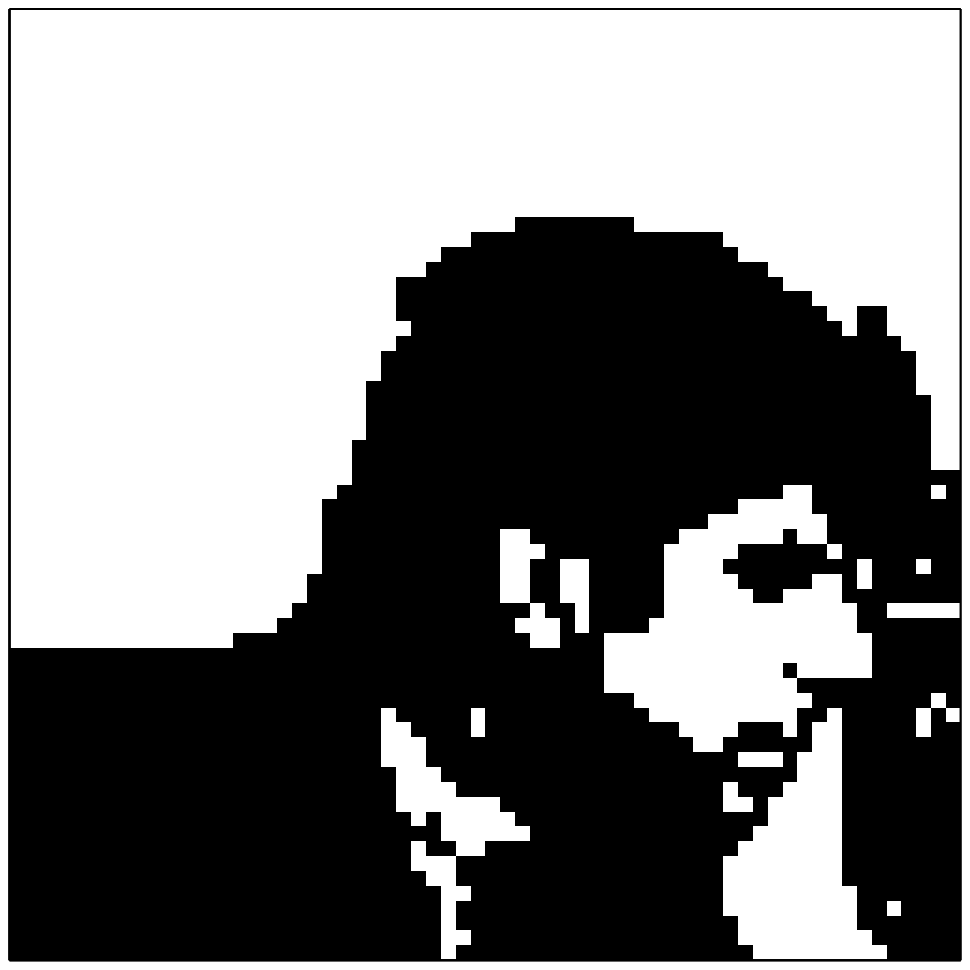}}}
      }
    \caption{Two $64 \times 64$ natural binary images: (a) chair, and (b) man.}
    \label{fig: chair man}
  \end{center}
\end{figure}

We also tested the performance of our concatenated MRF-ISI detector
on the two natural binary images ``chair'' and ``man''
shown in Fig.~\ref{fig: chair man}.  In these
images, the numbers of 0s and 1s are nearly equal, so we used the
equiprobable version of the MRF-ISI system.  

\begin{figure}[htbp]
\centerline{\epsfig{figure=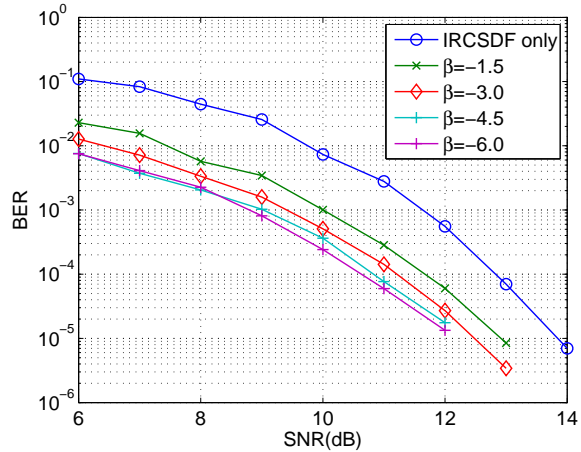, width=3.3in}}
\vspace{-0.1in}
\caption{Simulation results for the concatenated system of Fig.~\ref{fig: dblock}
on the 2D ISI channel with
$2 \times 2$ averaging mask and AWGN, on the natural binary image ``chair'' of
         Fig.~\ref{fig: chair man}, for various values of MRF parameter $\beta$.}
\label{chairsim}
\end{figure}

\begin{figure}[htbp]
\centerline{\epsfig{figure=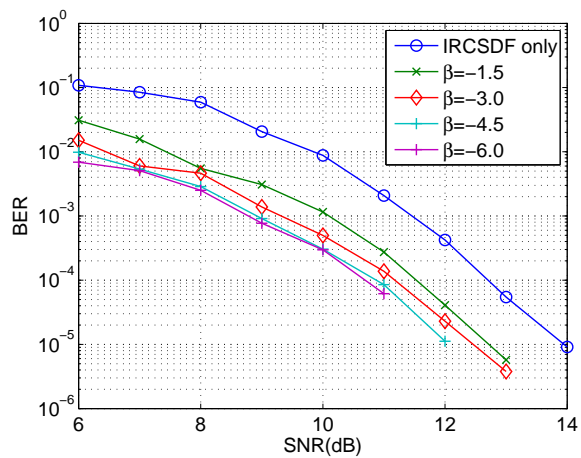, width=3.3in}}
\vspace{-0.1in}
\caption{Simulation results for the concatenated system of Fig.~\ref{fig: dblock}
on the 2D ISI channel with
$2 \times 2$ averaging mask and AWGN, on the natural binary image ``man'' of
         Fig.~\ref{fig: chair man}, for various values of MRF parameter $\beta$.}
\label{mansim}
\end{figure}

Without doing any model estimation, we simply used our first order MRF model with several 
guessed values of $\beta$ (namely, $-1.5$, $-3.0$, $-4.5$, and $-6.0$) to model these 
two images.  The simulation results for the
chair and man images appear in Figs.~\ref{chairsim} and \ref{mansim}.  For these
natural images at high SNR,
the concatenated detector achieves SNR savings of between 1 and 2 dB
compared to the ISI-only detector, for a wide choice of $\beta$ values at the receiver,
thereby demonstrating that a simple first-order MRF model is very useful in 
reducing 2D ISI in natural binary images.

\section{Conclusion}
\label{sec: conc}
This paper has demonstrated that, if the input image to a 2D ISI channel
has 2D correlation that can be modeled by a MRF, 
then significant SNR savings over previously proposed 2D-ISI detectors
can be realized by employing a concatenated iterative decoder consisting of 
SISO 2D-ISI and MRF detectors, and that the SNR savings increase with the degree of
source-image correlation.  The techniques described in this paper have
potential application in future-generation optical recording systems, 
which will employ 2D read/write heads.  In practice, many source images destined
for storage on such media are correlated; for example, 
uncompressed natural images are usually highly
correlated, and most practical image-compression
schemes leave residual correlation.  This paper has also demonstrated that
the proposed algorithm is quite robust to MRF parameter mismatch between the 
source image and the receiver, and that the simple first-order MRF is
a very useful model for 2D-ISI reduction in natural binary images.   
The basic ideas presented in this paper should
extend to more realistic image models and more practical scenarios.
Algorithms that learn an appropriately accurate MRF model from the source image 
would allow practical
implementation of joint 2D-ISI and MRF estimators in optical storage
systems; such algorithms are currently under investigation by the authors.

\section*{Acknowledgment}
% optional entry into table of contents (if used)
%\addcontentsline{toc}{section}{Acknowledgment}
The authors would like to thank the National
Science Foundation for providing partial support
for the work presented in this paper, under 
grants CCR-0098357 and CCF-0635390.



% http://www.ctan.org/tex-archive/macros/latex/contrib/supported/IEEEtran/bibtex
%\bibliographystyle{IEEEtran.bst}

\begin{thebibliography}{99}
\IEEEtriggeratref{12}

\bibitem{Geman}
S. Geman and D. Geman, ``Stochastic relaxation, {G}ibbs distributions, and the
{B}ayesian restoration of images,''
\emph{IEEE Trans.\ Pattern Analysis Mach.\ Intell.}, vol.\ PAMI-6, pp.~721--741, 
Nov.\ 1984.

\bibitem{in-phase-mit}
G. T. Huang, ``Holographic memory,'' {\em MIT Tech. Review}, vol. 9, Sep. 2005.
Available online at http://www.inphase-tech.com.

\bibitem{HGH96}
J.~F. Heanue, K.~G\"{u}rkan, and L.~Hesselink, ``Signal detection for
  page-access optical memories with intersymbol interference,'' {\em Applied
  Optics}, vol.~35, pp.~2431--2438, May 1996.

\bibitem{Kri98}
R.~Krishnamoorthi, ``Two-dimensional {Viterbi} like algorithms,'' Master's
  thesis, Univ. Illinois at Urbana Champaign, 1998.

\bibitem{CC98}
X.~Chen and K.~M. Chugg, ``Near-optimal data detection for two-dimensional
  {ISI/AWGN} channels using concatenated modeling and iterative algorithms,''
  in {\em Proc. IEEE International Conference on Communications, {ICC'98}},
  pp.~952--956, 1998.

\bibitem{MHNM97}
C. L. Miller, B. R. Hunt, M. W. Marcellin, and M. A. Neifeld, ``Image restoration 
using the Viterbi algorithm,'' {\em Journal of the Optical Society of America A},
Vol. 16, No. 2, pp. 265-274, Feb. 2000.

\bibitem{marrowconf03}
M.~Marrow and J.~K. Wolf, ``Iterative detection of 2-dimensional {ISI}
  channels,'' in {\em Proc. Info. Theory Workshop}, (Paris, France),
  pp.~131--134, Mar./Apr. 2003.

\bibitem{marrowvgs}
M.~Marrow, ``Equalization and detection of 2-d {ISI} channels.'' slide
  presentation available at http://cmrr-wolf08.ucsd.edu/\verb+~+mmarrow/, May
  2003.

\bibitem{OSullivan}
Y. Wu and J. A. O'Sullivan, ``Iterative Detection and Decoding for Separable Two-Dimensional 
Intersymbol Interference,'' {\em IEEE Trans. Magnetics},  Vol. 39, No.4, July 2003

\bibitem{shental} O. Shental, A. J. Weiss, N. Shental, and Y. Weiss, ``Generalized
belief propagation receiver for near-optimum detection of two-dimensional channels
with memory,'' {\em Proc. Info. Theory Workshop}, (San Antonio, Texas),
  pp.~225--229, Oct. 2004.

\bibitem{BL05}
Z.~Zhao and R.~E. Blahut, ``The Richardson-Lucy algorithm based demodulation algorithms for the two-dimensional intersymbol interference channel,'' in {\em Proc. 39th Conf. on Info. Sci. and 
Systems (CISS'05),} (CD-ROM), paper 83, Johns-Hopkins U., Mar. 2005.

\bibitem{taikun} 
T. Cheng, B. Belzer and K. Sivakumar, ``Image deblurring with iterative row-column 
soft-decision feedback algorithm,'' in {\em Proc. 39th Conf. on Info. Sci. and 
Systems (CISS'05),} (CD-ROM), paper 210, Johns-Hopkins U., Mar. 2005.

\bibitem{patrick} P. M. Njeim, T. Cheng, B. J. Belzer and K. Sivakumar, 
``Image detection in 2D intersymbol interference with iterative 
soft-decision feedback zig-zag algorithm,'' in {\em Proc. 43rd 
Allerton Conf. on Comm., Comp., and Control,} (CD-ROM), 
U. of Illinois Urbana-Champaign, Sept. 2005.

\bibitem{taikun_jour} T. Cheng, B. Belzer and K. Sivakumar, 
``An iterative row-column soft-decision feedback algorithm for 
two-dimensional intersymbol interference,'' submitted to {\em IEEE
Sig. Proc. Letters,} Nov. 2005.

\bibitem{Berrou}
C.~Berrou and A.~Glavieux, ``Near optimum error correcting coding and decoding:
  turbo-codes,'' {\em IEEE Trans. Commun.}, vol.~44, pp.~1261 -- 1271, Oct.
  1996.

\bibitem{bcjr}
L.~R. Bahl, J.~Cocke, F.~Jelinek, and J.~Raviv, ``Optimal decoding of linear
  codes for minimizing symbol error rate,'' {\em IEEE Trans. Info. Th.},
  vol.~20, pp.~284--287, Mar. 1974.

\bibitem{Cross}
G.~R. Cross and A.~K. Jain, ``Markov Random Field Texture Models,'' {\em IEEE Trans. Pattern Analysis and Machine Intelligence}, vol.~PAMI-5, pp.~25--39, Jan. 1983.

\bibitem{mertins1}
A. Mertins, ``Image recovery from noisy transmission using soft bits and Markov random field models,''
{\em Optical Engineering,} vol. 42, pp.~2893-2899, Oct. 2003.

\bibitem{mertins2}
J. Kliewer, N. Gortz, and A. Mertins, ``On iterative source-channel image decoding with {M}arkov random field source models,''
{\em Proc. IEEE ICASSP 2004,} vol. 4, pp.~661-664, May 2004.

\end{thebibliography}

\end{document}